\documentclass[referee]{aa} 

\usepackage{graphicx}
\usepackage{txfonts}
\usepackage{amsmath}
\usepackage{slashbox}

\begin{document}

   \title{Plasma environment effects on K lines of astrophysical interest}
   \subtitle{I. Atomic structure, radiative rates and Auger widths of oxygen ions}

   \author{J. Deprince
          \inst{1},
           M.~A. Bautista,
           \inst{2},
          S. Fritzsche
          \inst{3,4},
          J.~A. Garc\'ia
          \inst{5,6},
          T.~R. Kallman
          \inst{7},
          C. Mendoza
          \inst{2},
          P. Palmeri
          \inst{1}
          \and
          P. Quinet\inst{1,8}
          }

   \institute{Physique Atomique et Astrophysique, Universit\'e de Mons -- UMONS,
              B-7000 Mons, Belgium\\
              \email{patrick.palmeri@umons.ac.be}
         \and
              Department of Physics, Western Michigan University, Kalamazoo,
              MI 49008, USA
         \and
             Helmholtz Institut Jena, 07743 Jena, Germany
         \and
             Theoretisch Physikalisches Institut, Friedrich Schiller Universit\"at Jena,
             07743 Jena, Germany
         \and
             Cahill Center for Astronomy and Astrophysics, California Institute of Technology, Pasadena,
             CA 91125, USA
         \and
             Dr. Karl Remeis-Observatory and Erlangen Centre for Astroparticle Physics, Sternwartstr. 7, 
              96049 Bamberg, Germany
         \and
             NASA Goddard Space Flight Center, Code 662, Greenbelt, MD 20771, USA
         \and
             IPNAS, Universit\'e de Li\`ege, Sart Tilman, B-4000 Li\`ege, Belgium
             }

   \date{Received ??; accepted ??}


  \abstract
   {}
{In the context of black-hole accretion disks, the main goal of the present study is to estimate the plasma environment effects on the atomic structure and radiative parameters associated with the K-vacancy states in ions of the oxygen isonuclear sequence.}
{We use a time-averaged Debye--H\"uckel potential for both the electron--nucleus and the electron--electron interactions implemented in the fully relativistic multiconfiguration Dirac--Fock (MCDF) method.}
{Modified ionization potentials, K thresholds, Auger widths and radiative transition wavelengths and rates are reported for \ion{O}{i} -- \ion{O}{vii} in plasma environments with electron temperature and density ranges $10^5{-}10^7$~K and $10^{18}{-}10^{22}$~cm$^{-3}$.}
   {}


   \titlerunning{Plasma effects on oxygen K lines}

   \authorrunning{Deprince et al}

   \maketitle
%

\section{Introduction}

High-density plasma effects (free--free heating at electron densities $n_e > 10^{19}$~cm$^{-3}$) may explain the apparent supersolar Fe abundances inferred from the X-ray spectra of accreting black holes \citep{gar18}. However, the currently available atomic data to model astronomical spectra do not take into account high-density effects, and are therefore limited to densities below $10^{18}$~cm$^{-3}$ \citep{gar16}. In this respect \citet{sch13} recently carried out MHD simulations of a 10~M$_{\odot}$ black hole accreting at a 10\% rate, and predicted plasma conditions in the accretion disk characterized by electron temperatures and densities spanning the ranges $10^5{-}10^7$~K and $10^{18}{-}10^{22}$~cm$^{-3}$. \citet{fie07} have determined supersolar abundances for carbon, nitrogen, oxygen and iron from a high-resolution {\it Chandra} spectrum of the warm absorbing gas outflowing from the AGN Mrk 279 compact region, where especially the oxygen abundance is estimated at ${\sim}8$ times solar. This latter environment type is also expected to have densities greater than $10^{19}$~cm$^{-3}$.

\citet{gar05} reported a complete set of atomic data relevant to the modeling of oxygen K lines formed in astrophysical photoionized plasmas. As underlined by \citet{smi14} in the interpretation of X-ray spectra taken by current space observatories, none of these atomic parameters takes into account high-density plasma embedding, where the atomic structure and processes (e.g. emissivities, opacities and ionization balance) could be significantly distorted by the extreme temperature and density.

In the present study we provide a complete set of structure and radiative data for the modeling of oxygen K lines that include plasma environment effects. In Section~2, we describe the atomic physics computational approach, and in Section 3 we validate the time-averaged potential used to model the plasma environment by means of three test cases. In Section 4 we discuss in detail our results, and finally our conclusions are drawn in Section~5.

\section{Theoretical approach}

In the MCDF method \citep{gra80,mck80,gra88} the atomic state function (ASF) $\Psi$ is represented by a linear combination of configuration state functions (CSF) $\Phi$ of the same parity ($P$), total angular momentum ($J$) and angular-momentum  projection ($M$)
\begin{equation}
\Psi (\gamma, P, J, M) = \sum_i c_i~\Phi (\alpha_i, P, J, M)\ ,
\label{expan}
\end{equation}
where the CSFs are antisymmetrized products of orthonormal monoelectronic spin-orbitals of the form
\begin{equation}
\varphi_{n \kappa m} (r,\theta,\phi) = \frac{1}{r} \begin{pmatrix}
P_{n \kappa}(r)~\chi_{\kappa m}(\theta,\phi) \\
i~Q_{n \kappa} (r)~\chi_{-\kappa m}(\theta,\phi) \end{pmatrix}\ .
\label{spino}
\end{equation}
In Eq.~\eqref{spino} $P_{n \kappa}(r)$ and $Q_{n \kappa}(r)$ are the large and small radial orbitals, respectively, and the angular functions $\chi_{\kappa m}(\theta,\phi)$ are spinor spherical harmonics. These spin-orbitals are optimized self-consistently based on the Dirac--Coulomb Hamiltonian
\begin{equation}
H_{DC}=\sum_i c \vec{\alpha_i} \cdot \vec{p_i}+ \beta_i c^2 - \frac{Z}{r_i}
+ \sum_{i>j} \frac{1}{r_{ij}}\ .
\label{dc}
\end{equation}

For an atom embedded in a weakly coupled plasma, the Hamiltonian of Eq.~\eqref{dc} is replaced with the Debye-H\"uckel (DH) screened Dirac--Coulomb Hamiltonian \citep{sah06}
\begin{equation}
H^{DH}_{DC}=\sum_i c \vec{\alpha_i} \cdot \vec{p_i}+ \beta_i c^2 - \frac{Z}{r_i}
e^{-\mu r_i}
+ \sum_{i>j} \frac{1}{r_{ij}} e^{-\mu r_{ij}}\ ,
\label{dh}
\end{equation}
where $r_{ij}=|\vec{r}_i-\vec{r}_j|$ and the plasma screening parameter $\mu$ is the inverse of the Debye shielding length $\lambda_D$, which can be expressed in atomic units (a.u.) as a function of the plasma electron density $n_e$ and temperature $T_e$ as
\begin{equation}
\mu = \frac{1}{\lambda_D} = \sqrt{\frac{4\pi n_e}{k T_e}}\ .
\label{screen}
\end{equation}

Typical plasma conditions in black-hole accretion disks are $T_e\sim  10^5{-}10^7$~K and $n_e\sim 10^{18}{-}10^{22}$~cm$^{-3}$ \citep{sch13}. For weakly coupled plasmas they correspond to screening parameters $0.0\leq \mu\leq 0.24$~a.u. and, for a completely ionized hydrogen plasma (with plasma ionization $Z^*=1$), to plasma coupling parameters
\begin{equation}
\Gamma = \frac{e^2}{4\pi\epsilon_0 d kT_e}
\end{equation}
with
\begin{equation}
d=\left(\frac{3}{4\pi n_e}\right)^{1/3}\ ,
\end{equation}
in the range $0.0003\leq \Gamma \leq 0.6$.

The last term of Eq.~(\ref{dh}) has the angular dependence \citep{sah06}
\begin{equation}
\frac{1}{r_{ij}} e^{-\mu r_{ij}}
= - \mu \sum_{l=0}^{\infty}~(2l+1)~j_l(i \mu r_{<})
~h^1_l(i \mu r_{>})~P_l(cos~\omega_{ij})\ ,
\label{dhee}
\end{equation}
where $r_{>}=max(r_i,r_j)$, $r_{<}=min(r_i,r_j)$, $j_l$
is a Bessel function, $h^1_l$ denotes a Hankel function of the first kind and $P_l$ is a Legendre polynomial that depends on the angle $\omega_{ij}$ between the two position vectors $\vec{r}_i$ and $\vec{r}_j$. This screening reduces the electron--electron repulsion and, hence, increases the binding of the electron by the nucleus.

We use the active space (AS) method to obtain the MCDF expansions of Eq.~\eqref{expan} for \ion{O}{i} -- \ion{O}{vii}, whereby electrons from reference configurations are excited to a given active set of orbitals. For these oxygen ions, the AS was built up by considering all the single and double excitations of the reference configurations listed for each species in Table~\ref{as} to configurations including $n=2$ and $n=3$ orbitals.

\begin{table}[h!]
  \caption{Reference configurations used to build up the MCDF active space for \ion{O}{i} -- \ion{O}{vii}. \label{as}}
  \centering
  \small
  \begin{tabular}{ll}
  \hline\hline
  \noalign{\smallskip}
  Ion & Reference configurations \\
  \hline
  \noalign{\smallskip}
  \ion{O}{i}  & ${\rm 2p^4}$, ${\rm [2s]2p^5}$, ${\rm [2s]^22p^6}$, ${\rm [1s]2p^5}$, ${\rm [1s][2s]2p^6}$ \\
  \ion{O}{ii} & ${\rm 2p^3}$, ${\rm [2s]2p^4}$, ${\rm [2s]^22p^5}$, ${\rm [1s]2p^4}$, ${\rm [1s][2s]2p^5}$, ${\rm [1s][2s]^22p^6}$ \\
  \ion{O}{iii} & ${\rm 2p^2}$, ${\rm [2s]2p^3}$, ${\rm [2s]^22p^4}$, ${\rm [1s]2p^3}$, ${\rm [1s][2s]2p^4}$, ${\rm [1s][2s]^22p^5}$ \\
  \ion{O}{iv} & ${\rm 2p}$, ${\rm [2s]2p^2}$, ${\rm [2s]^22p^3}$, ${\rm [1s]2p^2}$, ${\rm [1s][2s]2p^3}$, ${\rm [1s][2s]^22p^4}$ \\
  \ion{O}{v}  & ${\rm 2s^2}$, ${\rm 2s2p}$, ${\rm 2p^2}$, ${\rm [1s]2s^22p}$, ${\rm [1s]2s2p^2}$, ${\rm [1s]2p^3}$ \\
  \ion{O}{vi} & ${\rm 2s}$, ${\rm 2p}$, ${\rm [1s]2s2p}$, ${\rm [1s]2p^2}$ \\
  \ion{O}{vii} & ${\rm 1s^2}$, ${\rm 1s2s}$, ${\rm 1s2p}$ \\
  \hline
  \end{tabular}
\end{table}

In the isolated atom approximation, the relativistic orbitals $P_{n\kappa}(r)$ and $Q_{n\kappa}(r)$, along with the expansion coefficients $c_i$ in Eqs.~\eqref{expan}--\eqref{spino}, were optimized using GRASP2K \citep{par96} with the extended average level (EAL) option, where the $(2J+1)$-weighted trace of the Dirac--Coulomb Hamiltonian (see Eq.~\ref{dc}) is minimized to determine energy levels, wavelengths and radiative and Auger rates. To take into account core-relaxation effects on the K-vacancy states \citep{gar05}, we introduce non-orthogonal orbitals optimized separately in two distinct level groups: a first group of exclusively valence levels where the K shell is full; and a second group of levels with at least a single K-vacancy. For the ionization potential (IP) and K threshold, the orbitals were respectively optimized on the ground level and on the lowest K-vacancy level of each ion using the optimal level (OL) option of GRASP2K.

Plasma effects are included perturbatively in a second step where we use the RATIP code of \citet{fri12} to improve the expansion coefficients, i.e. the ASF, energy levels and radiative rates by solving the secular equation with the Debye-H\"uckel screened Dirac--Coulomb Hamiltonian (Eq.~\ref{dh}), the two-body Breit interaction and the quantum electrodynamic corrections (self-energy and vacuum polarization). Plasma screening parameters in the range $0.00\leq\mu\leq 0.25$~a.u. were adopted, the upper-limit choice, justified in Section~3.3, corresponding to the extreme plasma conditions found in accretion disks.

\section{Validation of the Debye--H\"uckel model potential}

Stark shifts of the dipole-allowed spectral lines emitted by an ion in a dense plasma are measured. In a semi-classical picture, neighboring electrons and ions give rise to effective microscopic electric fields that result in level energy shifts due to induced dipole moments in the emitting ion. Alternatively, Stark shifts can also be predicted quantum mechanically with a Debye--H\"uckel potential \citep{roz75}. To test the validity of our Debye--H\"uckel model potential, we have performed a series of calculations with the GRASP2K/RATIP code to compare with available laboratory Stark shifts. Such comparisons are of common pratice, see e.g. \citet{roz75,nei76} and more recently \citet{bel15}. The three test cases investigated are detailed in the following Sections~\ref{stark1}--\ref{stark3}.

\subsection{Stark shifts in valence transitions of \ion{O}{ii}}
\label{stark1}

\citet{dje98} measured Stark shifts for the valence-shell transitions ${\rm 2p^23s~^4P_{3/2}}$ -- ${\rm 2p^23p~^4D^o_{5/2}}$ (${\lambda}4641.81$) and ${\rm 2p^23s~^4P_{1/2}}$ -- ${\rm 2p^23p~^4D^o_{3/2}}$ (${\lambda}4638.85$) in \ion{O}{ii}, respectively $0.03\pm 0.02$\,\AA\ and $0.05\pm 0.02$\,\AA, at a temperature $T_e=54\,000$\,K and density $n_e=2.8\times 10^{17}$\,cm$^{-3}$. To reproduce these shifts theoretically, we considered intravalence and core--valence correlations up to $n=5$ to represent the respective ASFs in the two transitions. The MCDF shifts obtained with $\mu = 0.0017$~a.u. are both 0.05 \AA\ in good agreement with experiment. Moreover, if the DH screening of the electron--electron Coulomb potential is switched off (i.e. $\mu = 0$ in the last term of Eq.~\ref{dh}), these shifts become much larger: 0.26 \AA\ and 0.25 \AA, respectively. This results confirm that the DH electron--electron plasma screening cannot be neglected. It should be also emphasied that
\citet{dje98} compared their measurements with the semi-classical calculations of \citet{gri74} and \citet{dim82} (see multiplet No.~1 in their Figure~4). The three sets of
Stark shifts \citep{dje98,gri74,dim82} disagree to each other, with the theoretical values having opposite signs and being about $\sim$ 0.15 \AA\ \citep{gri74} and $\sim-$0.05 \AA\ \citep{dim82}. In that respect, our time-independent quantum model confirms the measurements of \citet{dje98}.

\subsection{Stark shifts in valence transitions of \ion{Na}{i}}
\label{stark2}

The Stark shifts of the valence-shell \ion{Na}{i} D doublet at 5889.95/5895.90 \AA\ have been measured in a plasma with $T_e=38\,000$\,K and $n_e=3.5\times 10^{17}$\,cm$^{-3}$ \citep{sre96}: $0.38\pm 0.09$\,\AA\ and $0.41\pm 0.09$\,\AA, respectively. Our MCDF calculations with $\mu = 0.0023$~a.u. including both intravalence and core--valence correlations ($n\leq 4$) give 0.43\,\AA\ and 0.51\,\AA, respectively, in satisfactory agreement with experiment. Again, if the electron--electron screening is neglected, the MCDF shifts are way off at 4.42\,\AA\ and 4.50\,\AA.

\subsection{Ti K$\alpha$ line pressure shift}
\label{stark3}

Our last test case concerns the K$\alpha$ line ${\rm 1s^2~^1S_0 - 1s2p~^1P^o_1}$ of He-like \ion{Ti}{xxi} at 4749.73~eV. \citet{kha12} measured a line shift of $3.4\pm 1.0$\,eV, and inferred an electron temperature greater than ${\sim}3$\,keV and an electron density exceeding $10^{24}$\,cm$^{-3}$ from a hydrodynamic simulation of their laser-produced plasma. This experimental shift was then reproduced theoretically using an ion-sphere model by \citet{bel15} leading to an estimate of 3.4~eV at $T_e=3$~keV and $n_e = 4.2\times 10^{24}$~cm$^{-3}$. In the same plasma conditions adopted by \citet{bel15}, namely $\mu = 0.27$~a.u., our MCDF-DH model yields a line shift of 3.3 \,eV in good agreement with both the measurement and the ion-sphere calculation. We are therefore confident to consider plasma parameters up to $\mu = 0.27$~a.u.

\section{Results and discussion}

In the following Sections~\ref{ip}--\ref{kvanc} we examine the DH plasma screening effects on the oxygen K-line characteristics by considering screening parameters in the range $0\leq\mu \leq 0.25$~a.u., which, as shown in Table~\ref{conds}, can be associated to electron temperatures and densities in the ranges $10^5\le T_e\le 10^7$~K and $10^{18}\le n_e\le 10^{22}$~cm$^{-3}$.

\begin{table*}[t]
  \caption{Plasma screening parameter $\mu$ (a.u.) for different electron temperatures $T_e$ and densities $n_e$. \label{conds}}
  \centering
  \begin{tabular}{ c | c c c c c}
  \hline\hline
  \noalign{\smallskip}
  \backslashbox{$T_e$\,(K)}{$n_e$\,(cm$^{-3}$)} & $10^{18}$ & $10^{19}$ & $10^{20}$ & $10^{21}$ & $10^{22}$ \\
  \hline
  \noalign{\smallskip}
    $10^5$ & $\mu= 0.0024$ & $\mu= 0.0077$& $\mu= 0.024$& $\mu= 0.077$& $\mu= 0.24$\\
  \hline
  \noalign{\smallskip}
  $10^6$ & $\mu= 0.00077$ & $\mu= 0.0024$& $\mu= 0.0077$& $\mu= 0.024$& $\mu= 0.077$\\
  \hline
  \noalign{\smallskip}
  $10^7$ &$\mu= 0.00024$ & $\mu= 0.00077$& $\mu= 0.0024$& $\mu= 0.0077$& $\mu= 0.024$\\
  \hline
  \end{tabular}
\end{table*}

\subsection{Ionization potentials and K thresholds}
\label{ip}

The IPs ($E_0$) and K thresholds ($E_K$) for O ions determined with MCDF for three values of the plasma screening parameter---$\mu = 0$, 0.1 and 0.25~a.u---are listed in Tables~\ref{ipcomp}--\ref{kthcomp}. For isolated species ($\mu=0$), the NIST IPs \citep{nist} are reproduced to within 1\% except for \ion{O}{i}, for which only a 4\% accuracy was attained due to the well-known slow convergence of the CI expansion (Eq.~\ref{expan}) for neutrals. As expected, a substantial reduction of the absolute values of $E_0$ and $E_K$  increasing with $\mu$ is obtained. For the case of $\mu=0.1$ it is shown in Table~\ref{ipcomp} that, if the electron--electron DH screening is neglected ($\mu=0$ in the last term of Eq.~\ref{dh}), the continuum lowering, particularly for the lowly ionized species, is much larger yielding an unphysical negative IP for \ion{O}{i}. We therefore stress again the DH electron--electron screening must-have.

\begin{table}[h!]
  \caption{Plasma screening effects on the IP $E_0(\mu)$ in oxygen ions determined with the MCDF method. \label{ipcomp}}
  \centering
  \small
  \begin{tabular}{l c c c c c}
  \hline\hline
  \noalign{\smallskip}
  Ion & NIST$^a$ & \multicolumn{4}{c}{MCDF}\\
  \cline{3-6}
  \noalign{\smallskip}
  & $E_0$ & $E_0(0)$ & $E_0(0.1)$  & $E_0(0.1^b)$ & $E_0(0.25)$ \\
  \hline
  \noalign{\smallskip}
  \ion{O}{i}   & 13.61804(7)  & 13.07	& 10.33  & $-$6.85 & 6.42   \\
  \ion{O}{ii}  & 35.12111(6)  & 35.00   & 29.68  & 14.65   & 22.30  \\
  \ion{O}{iii} & 54.93554(12) & 54.80	& 46.90  & 34.26   & 36.12  \\
  \ion{O}{iv}  & 77.41350(25) & 77.31   & 66.82  & 56.60   & 52.52  \\
  \ion{O}{v}   & 113.8989(5)  & 112.81  & 99.75  & 92.04   & 82.14  \\
  \ion{O}{vi}  & 138.1189(21) &	138.04  & 122.41 & 117.21  & 101.37 \\
  \ion{O}{vii} & 739.32679(6) &	739.86  & 720.99 & 718.32  & 693.36 \\
  \hline
  \end{tabular}
  \tablefoot{IP $E_0$ is given in eV and the screening parameter $\mu$ in a.u.
  \tablefoottext{a}{\citet{nist}.}
  \tablefoottext{b}{e--e screening switched off.}
  }
\end{table}

\begin{table}[h!]
  \caption{Plasma screening effects on the K threshold $E_K(\mu)$ in oxygen ions computed with MCDF. \label{kthcomp}}
  \centering
  \small
  \begin{tabular}{l c c c}
  \hline\hline
  \noalign{\smallskip}
  Ion & $E_K(0)$ & $E_K(0.1)$ & $E_K(0.25)$ \\
  \hline
  \noalign{\smallskip}
  \ion{O}{i}   & 543.58 & 540.62 & 535.47 \\
  \ion{O}{ii}  & 570.89 & 565.31 & 556.49 \\
  \ion{O}{iii} & 593.27 & 585.08 & 572.74 \\
  \ion{O}{iv}  & 626.49 & 615.64 & 599.52 \\
  \ion{O}{v}   & 664.10 & 650.57 & 630.58 \\
  \ion{O}{vi}  & 699.64 & 683.44 & 659.64 \\
  \ion{O}{vii} & 739.86 & 720.99 & 693.36 \\
  \hline
  \end{tabular}
  \tablefoot{K threshold $E_K$ is given in eV and the screening parameter $\mu$ in a.u.}
\end{table}

\begin{figure}[!ht]
  \centering
  \includegraphics[width=0.9\columnwidth]{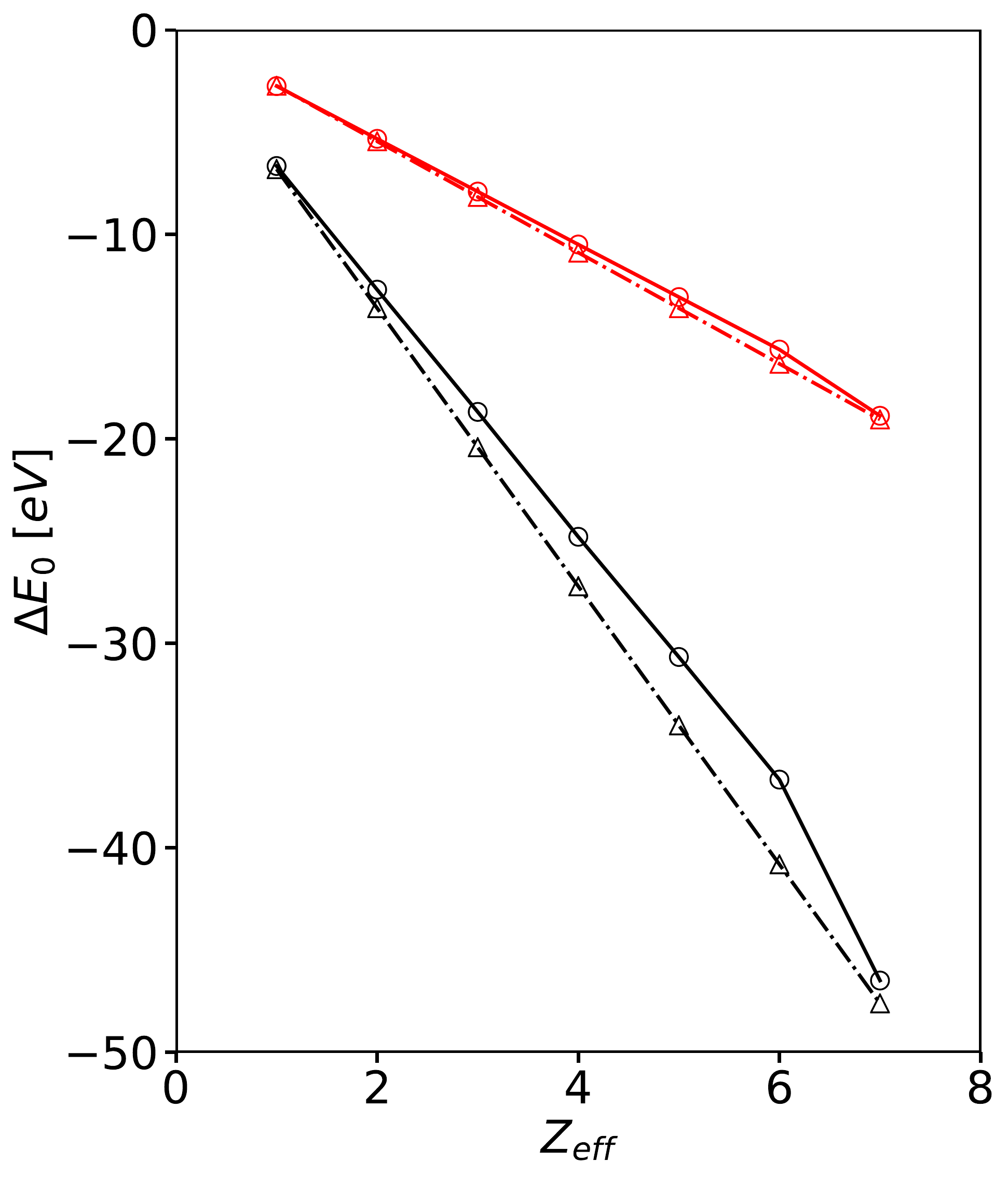}
  \caption{IP lowering $\Delta E_0 =E_0(\mu)-E_0(\mu=0)$ as function of the effective ionic charge $Z_{\rm eff}=Z-N+1$ for two different values of $\mu$. Circles: MCDF calculations. Triangles: DH limit ($\Delta E_0=-Z_{\rm eff}\,\mu$). Red: $\mu = 0.1$~a.u. Black: $\mu = 0.25$~a.u. The broken trends observed for $Z_{\rm eff}=7$ (\ion{O}{vii}) in MCDF calculations are due to the absence of electron--electron plasma screening in the ground state of \ion{O}{viii} that contributes to the ionization potentials of \ion{O}{vii}.}\label{ipvar}
\end{figure}

In Fig.~\ref{ipvar} we plot the trends of the IP lowering  $\Delta E_0(\mu) = E_0(\mu) - E_0(\mu=0)$ with $Z_{\rm eff}=Z-N+1$, which are found to be practically linear except for $Z_{\rm eff}=7$ (\ion{O}{vii}) due to the absence of the DH electron--electron screening. We also include for each species the DH limit $\Delta E^{\rm DH}_0 \equiv -Z_{\rm eff}\,\mu$  as $\Gamma \rightarrow 0$ determined by \citet{sp66} and \citet{crow14}. For \ion{O}{vii}, $\Delta E^{\rm DH}_0 =-47.6$\,eV close to the MCDF IP lowering for $\mu=0.25$.

\begin{figure}[!ht]
  \centering
  \includegraphics[width=0.9\columnwidth]{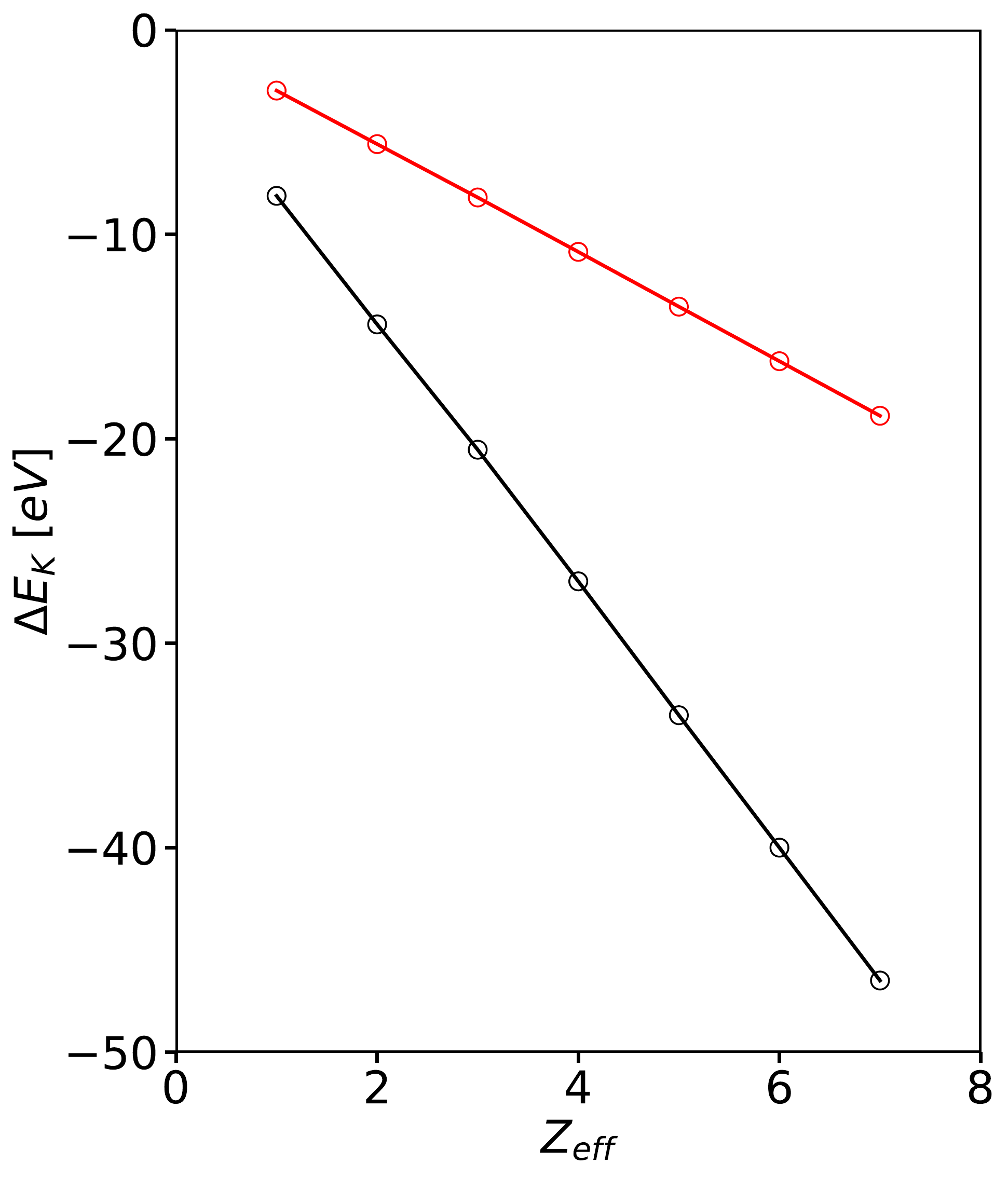}
  \caption{MCDF K-threshold lowering $\Delta E_K = E_K(\mu)-E_K(\mu=0)$ as function of the effective charge $Z_{\rm eff}=Z-N+1$ for two different values of $\mu$. Red: $\mu = 0.1$~a.u. Black: $\mu = 0.25$~a.u.} \label{kthvar}
\end{figure}

K-threshold lowering ($\Delta E_K(\mu) = E_K(\mu) - E_K (\mu=0)$) trends with $Z_{\rm eff}$ are very similar to those of $\Delta E_0$ as shown in Fig.~\ref{kthvar}, a predominantly linear decrease up to $\Delta E_K \approx\Delta E_0\approx -50$\,eV. This finding is significant inasmuch as the DH screened photoionization cross sections will only involve approximately constant downward energy shifts of the thresholds leading series truncations rather than variant line wavelengths and resonance energy positions, points that are  further discussed in Sections~\ref{rad}--\ref{kvanc}.

\subsection{Radiative data}
\label{rad}

Oxygen K-line wavelengths and transition probabilities ($A$-values) computed with MCDF with $\mu = 0$, 0.1 and 0.25~a.u. are reported in Table~\ref{lino}. For isolated systems ($\mu=0$) our radiative data are in good general accord with the HFR and MCBP results of \citet{gar05}; more precisely, present K-line wavelengths are shorter by 0.1\% for the highly charged ions to little less than 1\% for the lower ionization stages. In addition, they are in excellent agreement with the few spectroscopic reports available; for example, for the strong K$\alpha$ line in \ion{O}{vii}, our predicted wavelength agrees within 0.2\% with the measurement by \citet{eng95}, and within 0.7\%  and 0.2\% with those reported by \citet{sch04} for \ion{O}{v} and \ion{O}{vi}, respectively. Regarding radiative rates, our MCDF results agree with \citet{gar05} on average to within 10\% except for a few weak transitions.

\begin{figure}[!ht]
  \centering
  \includegraphics[width=0.7\columnwidth]{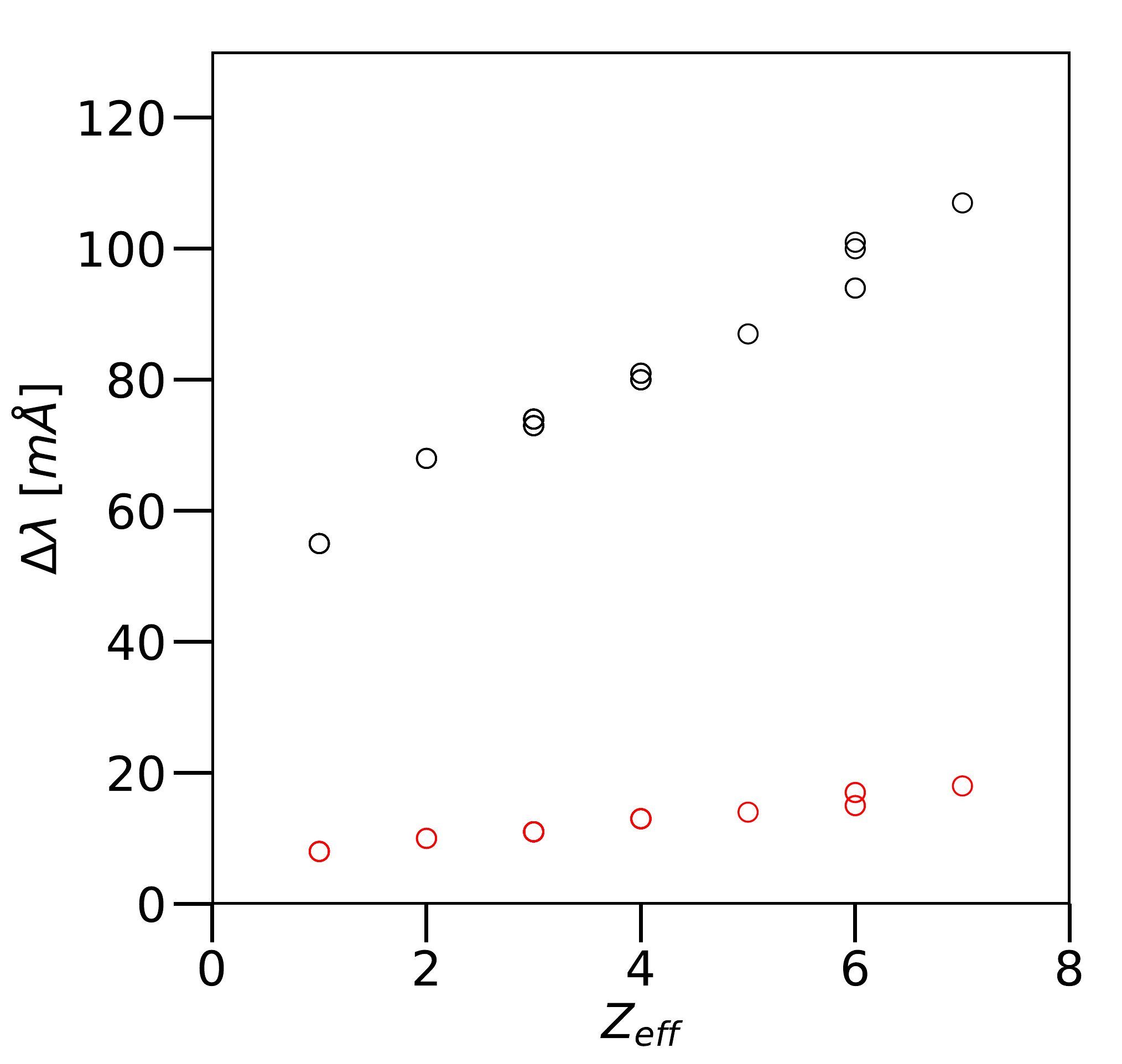}
  \includegraphics[width=0.7\columnwidth]{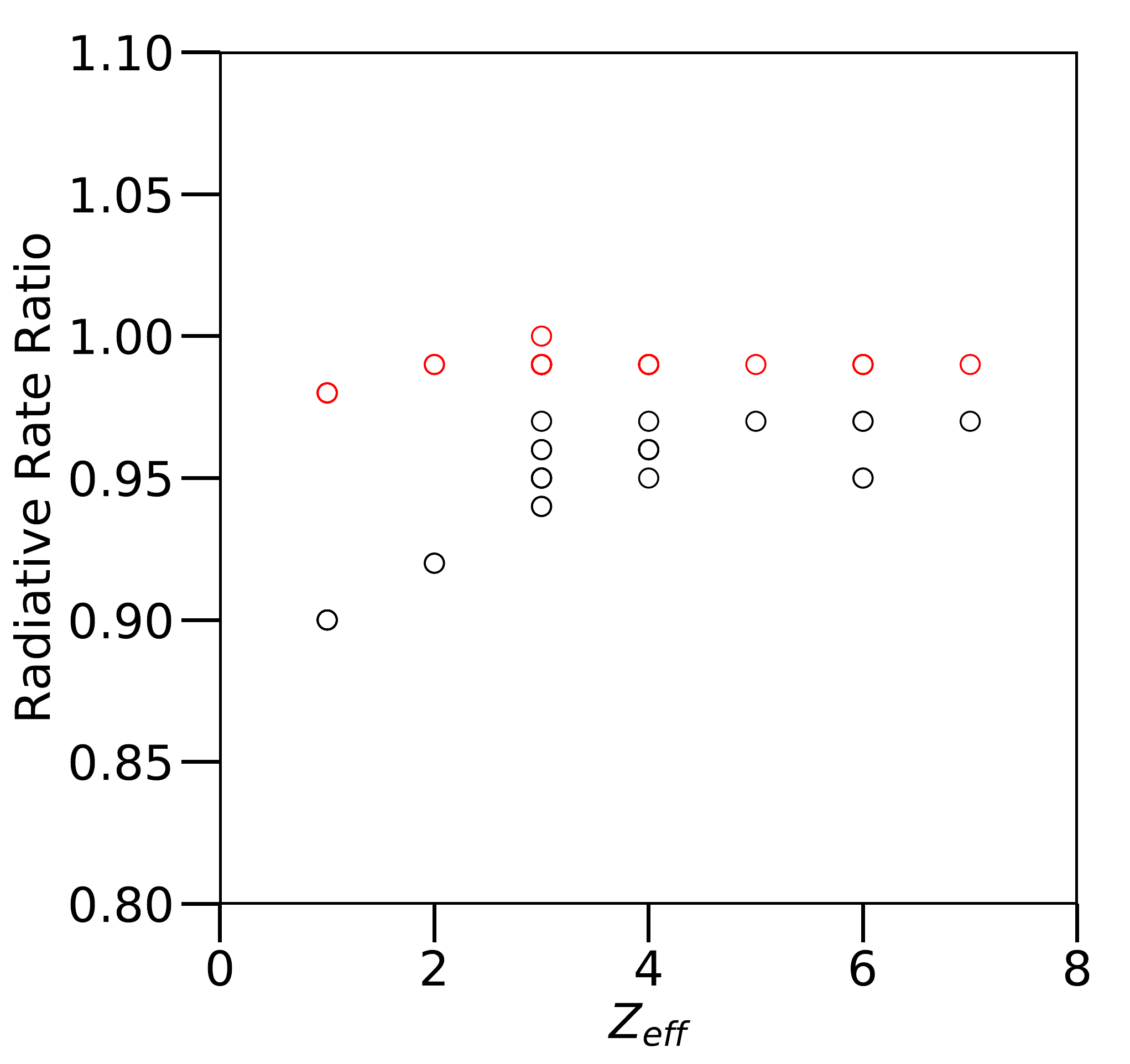}
  \caption{{\it Top}: MCDF wavelength pressure shift $\Delta\lambda = \lambda(\mu)- \lambda(\mu=0)$ for oxygen K lines as function of the effective charge $Z_{\rm eff}=Z-N+1$ for two different values of $\mu$. {\it Bottom}: MCDF radiative rate ratio $A(j,i,\mu)/A(j,i,\mu=0)$ for oxygen K lines as function of the effective charge $Z_{\rm eff}=Z-N+1$ for two different values of $\mu$. Red: $\mu = 0.1$~a.u. Black: $\mu = 0.25$~a.u.} \label{dwavel}
\end{figure}

Plasma effects on the radiative parameters are found to be small for $\mu= 0.1$~a.u. but more conspicuous for $\mu= 0.25$~a.u (see Fig.~\ref{dwavel}). In fact for $\mu= 0.25$~a.u. K-line wavelengths appear to be shifted by ${\sim}50{-}100$~m\AA\ with respect to the isolated atom ($\mu = 0$) with a trend increasing with the
ion effective charge $Z_{\rm eff}$ as shown in Fig.~\ref{dwavel} (upper panel). Although small, such wavelength shifts can be resolved by present and next-generation satellite borne X-ray spectrometers. $A$-values are generally also slightly modified; for instance, they are reduced on average by 5\% with $\mu= 0.25$~a.u. (see lower panel of Fig.~\ref{dwavel}), which would make negligible differences in astrophysical modeling.

\subsection{K-vacancy level energies and Auger widths}
\label{kvanc}

\begin{figure}[!ht]
  \centering
  \includegraphics[width=0.7\columnwidth]{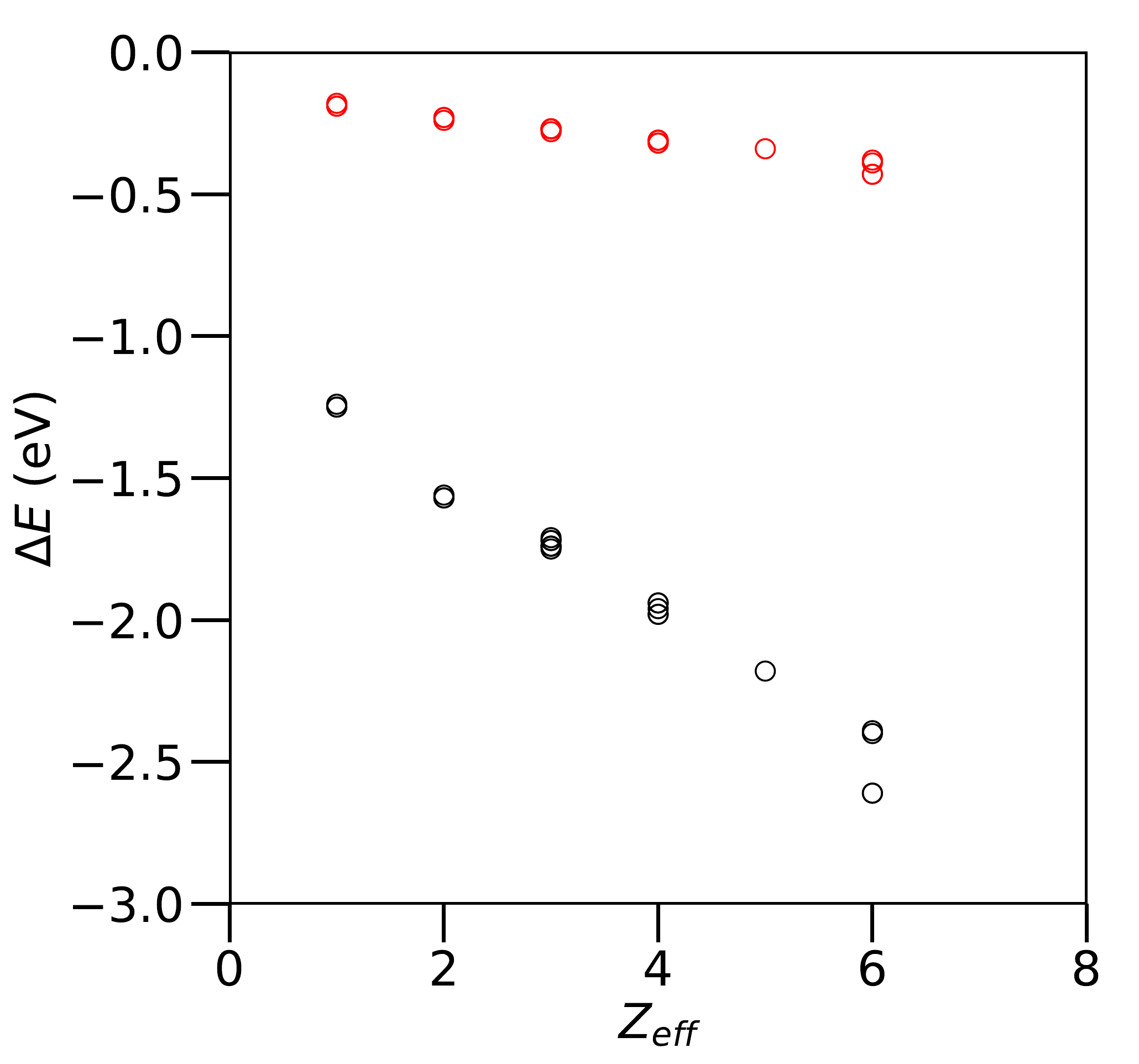}
  \includegraphics[width=0.7\columnwidth]{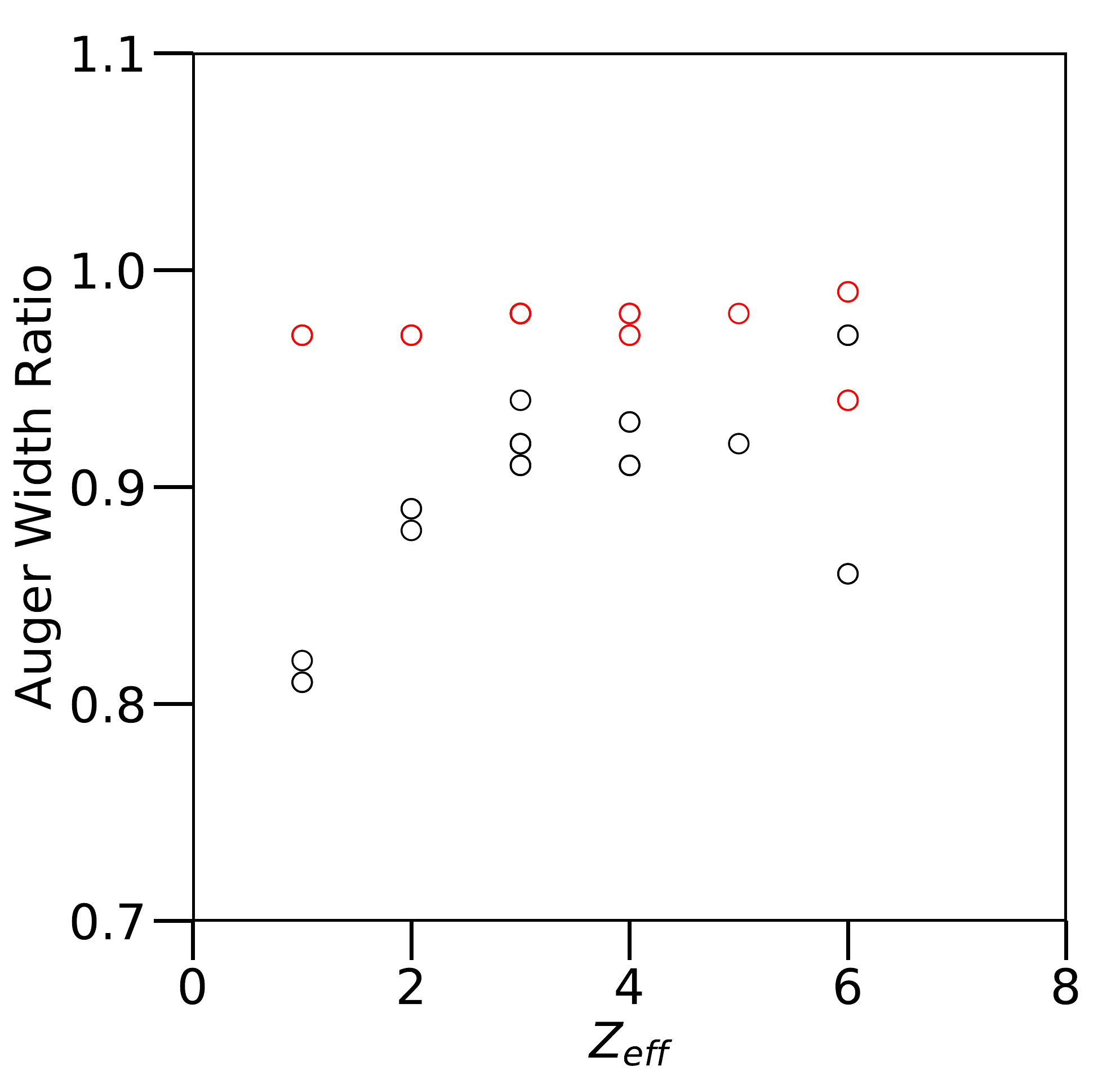}
  \caption{{\it Top}: MCDF level-energy pressure shift $\Delta E = E(\mu)-E(\mu=0)$ for oxygen K-vacancy levels as function of the effective charge $Z_{\rm eff}=Z-N+1$ for two different values of $\mu$. {\it Bottom}: MCDF Auger-width ratio $A_a(j,\mu)/A_a(j,\mu=0)$ for oxygen K-vacancy levels as function of the effective charge $Z_{\rm eff}=Z-N+1$ for two different values of $\mu$. Red: $\mu = 0.1$~a.u. Black: $\mu = 0.25$~a.u.} \label{dener}
\end{figure}

MCDF level energies and Auger widths for oxygen K-vacancy levels are presented for $\mu= 0$, 0.1 and 0.25~a.u. in Table~\ref{auger}. Our MCDF energies for $\mu=0$~a.u. are slightly lower ($\sim$0.5\% on average) with respect to those computed with the HFR and MCBP by \citet{gar05}, and our Auger widths are on average shorter by $\sim$25\% and $\sim$20\%, respectively.
Level-energy decrements are in general found to be small ($|\Delta E|\lesssim 3.0$\,eV) as illustrated in Fig.~\ref{dener} (top panel) for $\mu= 0.1$~a.u and $\mu= 0.25$~a.u. with trends increasing with $Z_{\rm eff}$. Regarding the Auger widths as shown in Fig.~\ref{dener} (bottom panel), they are reduced on average by up to $\sim 10\%$ for $\mu= 0.25$~a.u., which might have an impact on spectral K-line modeling. Neutral oxygen ($Z_{\rm eff}= 1$) has a different situation where MCDF predicts a 20 \% decrease for $\mu= 0.25$~a.u. This may illustrate the difficulty of our atomic structure model to compute accurate rates (better than 20\%) at the neutral end of an isonuclear sequence.

\section{Summary and conclusions}

We have studied plasma embedding effects on the atomic structure of oxygen ions, namely the K-shell radiative parameters and Auger widths, as a function of the screening parameter. Such plasma effects were modeled perturbatively in the MCDF framework with a time-independent DH potential. Our main findings and conclusions can be summarized as follows.

\begin{enumerate}
  \item The validity of our DH model has been benchmarked with Stark line-shift measurements \citep{sre96,dje98,kha12} for screening parameters as large as $\mu =0.27$~a.u. The latter value has been associated with the extreme density conditions found in accretion disks around compact objects \citep[see Table~\ref{conds} and ][]{gar16}. To obtain the desirable degree of agreement with experiment, the DH electron--electron screening must be taken into account.
 
  \item We have studied plasma screening effects for $\mu \leq 0.25$~a.u. finding considerable lowering (up to $\sim 50$~eV) of both the IPs and K thresholds. Such shifts could arguably enhance the ionization fractions and K-vacancy state populations or, at least, lead to erroneous spectral line identifications.

  \item Only a modest impact on the radiative and Auger data has been detected for $\mu \leq 0.1$~a.u., but it becomes more acute under the extreme plasma conditions of $\mu = 0.25$~a.u.: K-line wavelengths are systematically redshifted by up to ${\sim}0.1$~\AA\ with potential consequences on out/inflow velocity determinations; and Auger rates might decrease by up to $\sim 20\%$ in \ion{O}{i}.

  \item These new atomic data will be incorporated in the atomic the database of the XSTAR modeling code \citep{bau01} for future spectral analysis of accretion disks around compact objects.
      
  \item We believe the perturbative approach adopted here for $\mu\le 0.25$ is well supported by the relatively small effects on and smooth trends found in the atomic parameters, as well as by the good agreement with the experimentally determined Stark shifts for a few ionic species. More extreme conditions may require a non-perturbative inclusion of the DH potential; this work is underway and would be the subject of a subsequent report.
\end{enumerate}

\begin{acknowledgements}
JD is a Research Fellow of the Belgian Fund for Research in Industry and Agriculture FRIA. PP \& PQ are, respectively, Research Associate and Research Director of the Belgian Fund for Scientific Research F.R.S.--FNRS. Financial supports from these organizations, as well as from the NASA Astrophysics Research and Analysis Program (grant 80NSSC17K0345) are gratefully acknowledged. We are indented to Professor Nigel R. Badnell (Strathclyde University, UK) for lengthy and useful discussions on the validity the Debye--H\"uckel potential and its implementations in atomic structure calculations. JAG acknowledges support from the Alexander von Humboldt Foundation.
\end{acknowledgements}

%
%

\begin{table*}[t!]
  \caption{Plasma environment effects on the wavelengths and transition probabilities of K$\alpha$ lines in oxygen ions computed with MCDF. \label{lino}}
  \scriptsize
  \centering
  \begin{tabular}{l l c c c c c c c}
  \hline\hline
  \noalign{\smallskip}
Ion& \multicolumn{1}{c}{Transition} &\multicolumn{3}{c}{$\lambda$~(\AA)}&&\multicolumn{3}{c}{$A(j,i)$~(s$^{-1}$)}\\
  \cline{3-5}\cline{7-9}
  \noalign{\smallskip}
  &&$\mu=0.0$&$\mu=0.1$& $\mu=0.25$&&$\mu=0.0$&$\mu=0.1$ & $\mu=0.25$\\
  \hline
  \noalign{\smallskip}
  \ion{O}{i}  &$\left[\right.$1s$\left.\right]$2p$^5$ $^1$P$_1$ -- 2p$^4$ $^1$D$_2$	&	23.3180	&	23.3252	& 23.3684 &&	2.918E+12	&	2.885E+12&2.732E+12	\\
	&$\left[\right.$1s$\left.\right]$2p$^5$ $^3$P$_1$ -- 2p$^4$ $^3$P$_2$	&	23.3752	&	23.3832	& 23.4301 &&	6.648E+11	&	6.529E+11&5.993E+11	\\
	&$\left[\right.$1s$\left.\right]$2p$^5$ $^3$P$_0$ -- 2p$^4$ $^3$P$_1$	&	23.3753	&	23.3833	& 23.4302 &&	1.597E+12	&	1.568E+12&1.440E+12	\\
	&$\left[\right.$1s$\left.\right]$2p$^5$ $^3$P$_1$ -- 2p$^4$ $^3$P$_1$	&	23.3759	&	23.3839	& 23.4308 &&	3.992E+11	&	3.921E+11&3.601E+11	\\
	&$\left[\right.$1s$\left.\right]$2p$^5$ $^3$P$_1$ -- 2p$^4$ $^3$P$_0$	&	23.3763	&	23.3843	& 23.4311 &&	5.326E+11	&	5.231E+11&4.806E+11	\\
	&$\left[\right.$1s$\left.\right]$2p$^5$ $^3$P$_2$ -- 2p$^4$ $^3$P$_2$	&	23.3763	&	23.3843	& 23.4312 &&	1.197E+12	&	1.176E+12&1.080E+12	\\
	&$\left[\right.$1s$\left.\right]$2p$^5$ $^3$P$_2$ -- 2p$^4$ $^3$P$_1$	&	23.3770	&	23.3851	& 23.4319 &&	3.994E+11	&	3.924E+11&3.605E+11	\\
	&$\left[\right.$1s$\left.\right]$2p$^5$ $^1$P$_1$ -- 2p$^4$ $^1$S$_0$	&	23.3940	&	23.4001	& 23.4382 &&	5.923E+11	&	5.884E+11&5.704E+11	\\
  \hline
  \noalign{\smallskip}
  \ion{O}{ii}&$\left[\right.$1s$\left.\right]$2p$^4$ $^2$P$_{1/2}$ -- 2p$^3$ $^2$D$_{3/2}$	&	23.1144	&	23.1241	&23.1788 &&	2.534E+12	&	2.513E+12&2.407E+12	\\
	&$\left[\right.$1s$\left.\right]$2p$^4$ $^2$P$_{3/2}$ -- 2p$^3$ $^2$D$_{3/2}$	&	23.1159	&	23.1256	&23.1802 &&	2.541E+11	&	2.521E+11&2.426E+11	\\
	&$\left[\right.$1s$\left.\right]$2p$^4$ $^2$P$_{3/2}$ -- 2p$^3$ $^2$D$_{5/2}$	&	23.1159	&	23.1256	&23.1803 &&	2.342E+12	&   2.321E+12&2.224E+12 \\		
	&$\left[\right.$1s$\left.\right]$2p$^4$ $^2$S$_{1/2}$ -- 2p$^3$ $^2$P$_{1/2}$	&	23.1204	&	23.1303	&23.1858 &&	6.302E+11	&	6.231E+11&5.890E+11	\\
	&$\left[\right.$1s$\left.\right]$2p$^4$ $^2$S$_{1/2}$ -- 2p$^3$ $^2$P$_{3/2}$	&	23.1204	&	23.1303	&23.1858 &&	1.372E+12	&	1.358E+12&1.289E+12	\\
	&$\left[\right.$1s$\left.\right]$2p$^4$ $^2$D$_{5/2}$ -- 2p$^3$ $^2$D$_{3/2}$	&	23.1636	&	23.1734	&23.2286 &&	9.126E+10	&	9.033E+10&8.579E+10	\\
	&$\left[\right.$1s$\left.\right]$2p$^4$ $^2$D$_{5/2}$ -- 2p$^3$ $^2$D$_{5/2}$ &	23.1636	&	23.1734	&23.2287 &&	1.389E+12	&	1.375E+12&1.307E+12	\\
	&$\left[\right.$1s$\left.\right]$2p$^4$ $^2$D$_{3/2}$ -- 2p$^3$ $^2$D$_{3/2}$	&	23.1638	&	23.1736	&23.2288 &&	1.376E+12	&	1.362E+12&1.293E+12	\\
	&$\left[\right.$1s$\left.\right]$2p$^4$ $^2$D$_{3/2}$ -- 2p$^3$ $^2$D$_{5/2}$	&	23.1638	&	23.1736	&23.2289 &&	1.259E+11	&	1.247E+11&1.191E+11	\\
	&$\left[\right.$1s$\left.\right]$2p$^4$ $^2$P$_{1/2}$ -- 2p$^3$ $^2$P$_{1/2}$	&	23.1759	&	23.1852	&23.2376 &&	1.054E+12	&	1.047E+12&1.013E+12	\\
	&$\left[\right.$1s$\left.\right]$2p$^4$ $^2$P$_{1/2}$ -- 2p$^3$ $^2$P$_{3/2}$	&	23.1759	&	23.1852	&23.2376 &&	5.130E+11	&	5.093E+11&4.910E+11	\\
	&$\left[\right.$1s$\left.\right]$2p$^4$ $^2$P$_{3/2}$ -- 2p$^3$ $^2$P$_{1/2}$	&	23.1774	&	23.1867	&23.2390 &&	2.676E+11	&	2.657E+11&2.563E+11	\\
	&$\left[\right.$1s$\left.\right]$2p$^4$ $^2$P$_{3/2}$ -- 2p$^3$ $^2$P$_{3/2}$	&	23.1774	&	23.1867	&23.2391 &&	1.237E+12	&	1.229E+12&1.188E+12	\\
	&$\left[\right.$1s$\left.\right]$2p$^4$ $^4$P$_{1/2}$ -- 2p$^3$ $^4$S$_{3/2}$	&	23.2124	&	23.2227	&23.2807 &&	9.524E+11	&	9.400E+11&8.801E+11	\\
	&$\left[\right.$1s$\left.\right]$2p$^4$ $^4$P$_{3/2}$ -- 2p$^3$ $^4$S$_{3/2}$	&	23.2131	&	23.2234	&23.2814 &&	9.523E+11	&	9.399E+11&8.801E+11	\\
	&$\left[\right.$1s$\left.\right]$2p$^4$ $^4$P$_{5/2}$ -- 2p$^3$ $^4$S$_{3/2}$	&	23.2143	&	23.2246	&23.2825 &&	9.521E+11	&	9.397E+11&8.800E+11	\\
	&$\left[\right.$1s$\left.\right]$2p$^4$ $^2$D$_{5/2}$ -- 2p$^3$ $^2$P$_{3/2}$	&	23.2254	&	23.2348	&23.2877 &&	5.013E+11	&	4.973E+11&4.781E+11	\\
	&$\left[\right.$1s$\left.\right]$2p$^4$ $^2$D$_{3/2}$ -- 2p$^3$ $^2$P$_{1/2}$	&	23.2256	&	23.2350	&23.2879 &&	3.980E+11	&	3.949E+11&3.800E+11	\\
	&$\left[\right.$1s$\left.\right]$2p$^4$ $^2$D$_{3/2}$ -- 2p$^3$ $^2$P$_{3/2}$	&	23.2256	&	23.2350	&23.2879 &&	8.319E+10	&	8.247E+10&7.888E+10	\\
  \hline
  \noalign{\smallskip}
  \ion{O}{iii} &$\left[\right.$1s$\left.\right]$2p$^3$ $^1$P$_1$ -- 2p$^2$ $^1$D$_2$	&	22.8154	&	22.8267	&22.8887 &&	1.982E+12	&	1.967E+12&1.890E+12	\\
		&$\left[\right.$1s$\left.\right]$2p$^3$ $^3$S$_1$ -- 2p$^2$ $^3$P$_0$	&	22.8748	&	22.8862	&22.9485 &&	4.582E+11	&	4.554E+11&4.428E+11	\\
		&$\left[\right.$1s$\left.\right]$2p$^3$ $^3$S$_1$ -- 2p$^2$ $^3$P$_1$ &	22.8754	&	22.8868	&22.9491 &&	1.458E+12	&	1.449E+12&1.405E+12	\\
		&$\left[\right.$1s$\left.\right]$2p$^3$ $^3$S$_1$ -- 2p$^2$ $^3$P$_2$	&	22.8766	&	22.8880	&22.9503 &&	2.736E+12	&	2.717E+12&2.621E+12	\\
		&$\left[\right.$1s$\left.\right]$2p$^3$ $^3$P$_1$ -- 2p$^2$ $^3$P$_0$ &	22.8897	&	22.9011	&22.9637 &&	4.266E+11	&	4.225E+11&4.011E+11	\\
		&$\left[\right.$1s$\left.\right]$2p$^3$ $^3$P$_2$ -- 2p$^2$ $^3$P$_1$	&	22.8902	&	22.9016	&22.9642 &&	2.660E+11	&	2.636E+11&2.515E+11	\\
		&$\left[\right.$1s$\left.\right]$2p$^3$ $^3$P$_0$ -- 2p$^2$ $^3$P$_1$	&	22.8903	&	22.9017	&22.9642 &&	1.143E+12	&	1.132E+12&1.079E+12	\\
		&$\left[\right.$1s$\left.\right]$2p$^3$ $^3$P$_1$ -- 2p$^2$ $^3$P$_1$	&	22.8903	&	22.9017	&22.9643 &&	3.956E+11	&	3.918E+11&3.702E+11	\\
		&$\left[\right.$1s$\left.\right]$2p$^3$ $^1$D$_2$ -- 2p$^2$ $^1$D$_2$	&	22.8908	&	22.9021	&22.9636 &&	3.479E+12	&	3.455E+12&3.340E+12	\\
		&$\left[\right.$1s$\left.\right]$2p$^3$ $^3$P$_2$ -- 2p$^2$ $^3$P$_2$	&	22.8914	&	22.9028	&22.9653 &&	8.768E+11	&	8.684E+11&8.277E+11	\\
		&$\left[\right.$1s$\left.\right]$2p$^3$ $^3$P$_1$ -- 2p$^2$ $^3$P$_2$	&	22.8915	&	22.9029	&22.9654 &&	3.373E+11	&	3.342E+11&3.225E+11	\\
		&$\left[\right.$1s$\left.\right]$2p$^3$ $^1$P$_1$ -- 2p$^2$ $^1$S$_0$	&	22.9227	&	22.9338	&22.9943 &&	1.535E+12	&	1.525E+12&1.480E+12	\\
		&$\left[\right.$1s$\left.\right]$2p$^3$ $^3$D$_1$ -- 2p$^2$ $^3$P$_0$	&	22.9663	&	22.9776	&23.0395 &&	6.370E+11	&	6.319E+11&6.066E+11	\\
		&$\left[\right.$1s$\left.\right]$2p$^3$ $^3$D$_2$ -- 2p$^2$ $^3$P$_1$	&	22.9669	&	22.9782	&23.0400 &&	8.592E+11	&	8.522E+11&8.181E+11	\\
		&$\left[\right.$1s$\left.\right]$2p$^3$ $^3$D$_1$ -- 2p$^2$ $^3$P$_1$	&	22.9669	&	22.9782	&23.0401 &&	4.554E+11	&	4.518E+11&4.341E+11	\\
		&$\left[\right.$1s$\left.\right]$2p$^3$ $^3$D$_3$ -- 2p$^2$ $^3$P$_2$	&	22.9680	&	22.9792	&23.0410 &&	1.119E+12	&	1.110E+12&1.067E+12	\\
		&$\left[\right.$1s$\left.\right]$2p$^3$ $^3$D$_2$ -- 2p$^2$ $^3$P$_2$	&	22.9681	&	22.9794	&23.0412 &&	2.607E+11	&	2.587E+11&2.488E+11	\\
		&$\left[\right.$1s$\left.\right]$2p$^3$ $^3$D$_1$ -- 2p$^2$ $^3$P$_2$	&	22.9682	&	22.9794	&23.0413 &&	2.776E+10	&	2.755E+10&2.651E+10	\\
  \hline
  \noalign{\smallskip}
 \ion{O}{iv}&$\left[\right.$1s$\left.\right]$2p$^2$ $^2$S$_{1/2}$ -- 2p $^2$P$_{1/2}$	&	22.5066	&	22.5192	&22.5869 &&	3.799E+11	&	3.772E+11&3.636E+11	\\
		&$\left[\right.$1s$\left.\right]$2p$^2$ $^2$S$_{1/2}$ -- 2p $^2$P$_{3/2}$ &	22.5087	&	22.5214	&22.5890 &&	9.693E+11	&	9.607E+11&9.194E+11	\\
		&$\left[\right.$1s$\left.\right]$2p$^2$ $^2$P$_{3/2}$ -- 2p $^2$P$_{1/2}$	&	22.5567	&	22.5695	&22.6379 &&	6.296E+11	&	6.259E+11&6.072E+11	\\
		&$\left[\right.$1s$\left.\right]$2p$^2$ $^2$P$_{1/2}$ -- 2p $^2$P$_{1/2}$	&	22.5587	&	22.5716	&22.6399 &&	2.703E+12	&	2.686E+12&2.602E+12	\\
		&$\left[\right.$1s$\left.\right]$2p$^2$ $^2$P$_{3/2}$ -- 2p $^2$P$_{3/2}$	&	22.5588	&	22.5717	&22.6400 &&	3.323E+12	&	3.303E+12&3.203E+12	\\
		&$\left[\right.$1s$\left.\right]$2p$^2$ $^2$P$_{1/2}$ -- 2p $^2$P$_{3/2}$	&	22.5609	&	22.5737	&22.6420 &&	1.246E+12	&	1.239E+12&1.204E+12	\\
		&$\left[\right.$1s$\left.\right]$2p$^2$ $^2$D$_{3/2}$ -- 2p $^2$P$_{1/2}$	&	22.6310	&	22.6437	&22.7113 &&	1.102E+12	&	1.094E+12&1.058E+12	\\
		&$\left[\right.$1s$\left.\right]$2p$^2$ $^2$D$_{5/2}$ -- 2p $^2$P$_{3/2}$	&	22.6330	&	22.6457	&22.7132 &&	1.286E+12	&	1.277E+12&1.235E+12	\\
		&$\left[\right.$1s$\left.\right]$2p$^2$ $^2$D$_{3/2}$ -- 2p $^2$P$_{3/2}$	&	22.6331	&	22.6458	&22.7134 &&	1.862E+11	&	1.850E+11&1.791E+11	\\
  \hline
  \noalign{\smallskip}
  \ion{O}{v}  &$\left[\right.$1s$\left.\right]$2p $^1$P$_{1}$ -- 2s$^2$ $^1$S$_0$	&	22.2088	&	22.2229	&22.2960 &&	2.884E+12	&	2.868E+12&2.786E+12	\\
  \hline
  \noalign{\smallskip}
  \ion{O}{vi} &$\left[\right.$1s$\left.\right]$2s2p $^2$P$_{3/2}$ -- 2s $^2$S$_{1/2}$	&	21.7832	&	21.7998	&21.8840 &&	6.424E+11	&	6.368E+11&6.109E+11	\\	
		&$\left[\right.$1s$\left.\right]$2s2p $^2$P$_{1/2}$ -- 2s $^2$S$_{1/2}$	&	21.7836	&	21.8002	&21.8844 &&	6.792E+11	&	6.735E+11&6.467E+11	\\
		&$\left[\right.$1s$\left.\right]$2s2p $^2$P$_{3/2}$ -- 2s $^2$S$_{1/2}$	&	21.9706	&	21.9860	&22.0646 &&	2.694E+12	&	2.680E+12&2.609E+12	\\
		&$\left[\right.$1s$\left.\right]$2s2p $^2$P$_{1/2}$ -- 2s $^2$S$_{1/2}$	&	21.9730	&	21.9884	&22.0669 &&	2.657E+12	&	2.643E+12&2.573E+12	\\
  \hline
  \noalign{\smallskip}
  \ion{O}{vii}&1s2p $^1$P$_1$ -- 1s$^2$ $^1$S$_0$	&	21.5642	&	21.5821	&21.6707 &&	3.702E+12	&	3.680E+12&3.574E+12	\\
  \hline
  \end{tabular}
  \tablefoot{The plasma screening parameter $\mu$ is given in a.u., $\mu=0$ denoting the isolated atomic system.}
\end{table*}

\begin{table*}[t!]
  \caption{Plasma environment effects on the energy and Auger widths of K-vacancy states in oxygen ions computed with MCDF. \label{auger}}
  \scriptsize
  \centering
\begin{tabular}{l l c c c c c c c}
  \hline\hline
  \noalign{\smallskip}
  Ion& \multicolumn{1}{c}{Level} &\multicolumn{3}{c}{$E$~(eV)}&&\multicolumn{3}{c}{$A_a(j)$~(s$^{-1}$)}\\
  \cline{3-5}\cline{7-9}
  \noalign{\smallskip}
  &&$\mu=0.0$&$\mu=0.1$& $\mu=0.25$&&$\mu=0.0$&$\mu=0.1$ & $\mu=0.25$\\
  \hline
  \noalign{\smallskip}
  \ion{O}{i}&$\left[\right.$1s$\left.\right]$2p$^5$ $^3$P$_2$ &530.39	&530.20	&529.14 &&	2.606E+14	& 2.528E+14 & 2.104E+14	\\
	&$\left[\right.$1s$\left.\right]$2p$^5$ $^3$P$_1$ &530.41	&530.23	&529.17 &&	2.613E+14	& 2.536E+14	& 2.124E+14	\\
	&$\left[\right.$1s$\left.\right]$2p$^5$ $^3$P$_0$ &530.43	&530.24	&529.18 &&	2.587E+14	& 2.513E+14	& 2.110E+14	\\
	&$\left[\right.$1s$\left.\right]$2p$^5$ $^1$P$_1$ &534.13	&533.94	&532.87 &&	2.086E+14	& 2.037E+14	& 1.832E+14	\\
  \hline
  \noalign{\smallskip}
  \ion{O}{ii}&$\left[\right.$1s$\left.\right]$2p$^4$ $^4$P$_{5/2}$	&534.08	&533.85 &532.52 && 2.489E+14	& 2.405E+14 &  2.202E+14	\\
	&$\left[\right.$1s$\left.\right]$2p$^4$ $^4$P$_{3/2}$	&534.11	&533.87	&532.54 && 2.474E+14	& 2.392E+14 &  2.191E+14	\\
	&$\left[\right.$1s$\left.\right]$2p$^4$ $^4$P$_{1/2}$	&534.12	&533.89	&532.55 && 2.475E+14	& 2.394E+14 &  2.194E+14	\\
	&$\left[\right.$1s$\left.\right]$2p$^4$ $^2$D$_{3/2}$	&539.22	&538.96	&537.51 &&3.004E+14	& 2.928E+14 &  2.680E+14	\\
	&$\left[\right.$1s$\left.\right]$2p$^4$ $^2$D$_{5/2}$	&539.22	&538.96	&537.51 &&3.011E+14	& 2.935E+14 &  2.686E+14	\\
	&$\left[\right.$1s$\left.\right]$2p$^4$ $^2$P$_{3/2}$	&540.33	&540.07	&538.63 &&2.050E+14	& 1.975E+14 &  1.804E+14	\\
	&$\left[\right.$1s$\left.\right]$2p$^4$ $^2$P$_{1/2}$	&540.37	&540.11	&538.66 &&2.041E+14	& 1.969E+14 & 1.798E+14	\\
	&$\left[\right.$1s$\left.\right]$2p$^4$ $^2$S$_{1/2}$	&541.65	&541.38	&539.85 &&3.014E+14	& 2.924E+14 & 2.578E+14	\\
  \hline
  \noalign{\smallskip}
  \ion{O}{iii}&$\left[\right.$1s$\left.\right]$2p$^3$ $^3$D$_1$	&539.86	&539.59	&538.14 && 2.969E+14	& 2.903E+14	&2.719E+14	\\	
		&$\left[\right.$1s$\left.\right]$2p$^3$ $^3$D$_2$	&539.86	&539.59	&538.14 &&2.957E+14	& 2.891E+14	&2.709E+14	\\	
		&$\left[\right.$1s$\left.\right]$2p$^3$ $^3$D$_3$	&539.86	&539.59	&538.15 &&2.974E+14	& 2.908E+14	&2.724E+14	\\		
		&$\left[\right.$1s$\left.\right]$2p$^3$ $^3$P$_1$	&541.66	&541.39	&539.92 &&2.746E+14	& 2.682E+14	&2.500E+14	\\	
		&$\left[\right.$1s$\left.\right]$2p$^3$ $^3$P$_0$	&541.66	&541.39	&539.92 &&2.733E+14	& 2.669E+14	&2.488E+14	\\	
		&$\left[\right.$1s$\left.\right]$2p$^3$ $^3$P$_2$	&541.66	&541.39	&539.92 &&2.746E+14	& 2.681E+14	&2.499E+14	\\	
		&$\left[\right.$1s$\left.\right]$2p$^3$ $^3$S$_1$	&542.02	&541.74	&540.27 &&1.050E+14	& 1.031E+14	&9.827E+13	\\		
		&$\left[\right.$1s$\left.\right]$2p$^3$ $^1$D$_2$	&544.59	&544.30	&542.76 &&2.562E+14	& 2.505E+14	&2.350E+14	\\	
		&$\left[\right.$1s$\left.\right]$2p$^3$ $^1$P$_1$	&546.38	&546.09	&544.52 &&2.389E+14	& 2.333E+14	&2.177E+14	\\
  \hline
  \noalign{\smallskip}
  \ion{O}{iv} &$\left[\right.$1s$\left.\right]$2p$^2$ $^2$D$_{3/2}$	&547.85	&547.54& 545.91	&& 2.617E+14	& 2.552E+14	&2.392E+14	\\
		&$\left[\right.$1s$\left.\right]$2p$^2$ $^2$D$_{5/2}$	&547.86	&547.55&545.92	&& 2.622E+14	& 2.556E+14	&2.395E+14	\\
		&$\left[\right.$1s$\left.\right]$2p$^2$ $^2$P$_{1/2}$	&549.61	&549.29&547.63	&& 1.054E+14	& 1.033E+14	&9.766E+13	\\
		&$\left[\right.$1s$\left.\right]$2p$^2$ $^2$P$_{3/2}$	&549.66	&549.34&547.68	&& 1.054E+14	& 1.032E+14	&9.762E+13	\\	
		&$\left[\right.$1s$\left.\right]$2p$^2$ $^2$S$_{1/2}$	&550.88	&550.57&548.92	&& 2.239E+14	& 2.179E+14	&2.029E+14	\\
  \hline
  \noalign{\smallskip}
  \ion{O}{v}  &$\left[\right.$1s$\left.\right]$2p $^1$P$_{1}$	&558.26	&557.92	& 556.08&& 1.196E+14	& 1.168E+14 & 1.095E+14 \\
  \hline
  \noalign{\smallskip}
  \ion{O}{vi} &$\left[\right.$1s$\left.\right]$2s2p $^2$P$_{1/2}$	&564.25	&563.87&561.86	&& 3.902E+13	& 3.677E+13 &3.360E+13	\\
		&$\left[\right.$1s$\left.\right]$2s2p $^2$P$_{3/2}$	&564.32	&563.93&561.92	&& 3.651E+13	& 3.434E+13 & 3.128E+13	\\
		&$\left[\right.$1s$\left.\right]$2s2p $^2$P$_{1/2}$	&569.16	&568.73&566.55	&& 9.593E+13	& 9.544E+13 &9.314E+13	\\
		&$\left[\right.$1s$\left.\right]$2s2p $^2$P$_{3/2}$	&569.17	&568.74&566.56	&& 9.701E+13	& 9.649E+13 & 9.414E+13	\\
  \hline
  \end{tabular}
  \tablefoot{The plasma screening parameter $\mu$ is given in a.u., $\mu=0$ denoting the isolated atomic system.
  }
\end{table*}

\end{document}